\documentclass[a4paper,11pt]{article}
\usepackage{siunitx}
\usepackage{pos}
\usepackage{tabularx}
\usepackage{makecell}
\usepackage{ragged2e}
\usepackage{lipsum} 
\usepackage[mathlines]{lineno}

\title{Measurement of the Diffuse Astrophysical Neutrino Spectrum above a TeV with All Flavor Starting Events in IceCube}

\ShortTitle{IceCube All Flavor Diffuse Spectrum}

\author{The IceCube Collaboration \\{\normalsize \normalfont(a complete list of authors can be found at the end of the proceedings)}\\}

\emailAdd{vedant.basu@icecube.wisc.edu}
\emailAdd{aswathi.balagopalv@icecube.wisc.edu}
\emailAdd{karle@icecube.wisc.edu}

\abstract{

The IceCube Neutrino Observatory utilizes the Cherenkov radiation emitted by charged secondary particles produced in interactions of neutrinos with ice nucleons to detect neutrino events. “Starting events”, where this interaction vertex is contained inside the detector volume, can be used to distinguish neutrinos from the dominant background of atmospheric through-going muons. We present the Medium Energy Starting Events (MESE) selection, which employs a series of vetoes to obtain a neutrino-pure sample to measure the flux of diffuse extragalactic neutrinos from 1~TeV to 10~PeV from the entire sky. In this talk we will present a measurement of the spectrum of the diffuse flux of neutrinos, which demonstrates strong evidence for structure in the spectrum beyond a single power law, with a significance of $4.2\,\sigma$.

\vspace{4mm}

{\bfseries Corresponding authors:}
Vedant Basu$^{1*}$, 
Aswathi Balagopal V.$^{2}$, 
Albrecht Karle$^{3}$, 
\\
{$^{1}$ \itshape University of Utah}\\
{$^{2}$ \itshape University of Delaware}\\
{$^{3}$ \itshape University of Wisconsin-Madison}\\[4mm]
$^*$ Presenter
}

\ConferenceLogo{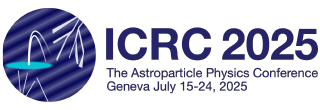}

\FullConference{39th International Cosmic Ray Conference (ICRC2025)\\
 15–24 July 2025\\
Geneva, Switzerland\\}

\begin{document}

\maketitle

\section{Introduction}\label{Introduction}

The IceCube Neutrino Observatory is a cubic-kilometer ice Cherenkov telescope located at the South Pole. The detector uses the clear Antarctic ice as a Cherenkov medium to detect high-energy astrophysical neutrinos through interactions with nucleons in the ice. At energies above 1~TeV, the dominant interaction mode is deep inelastic scattering (DIS). The Cherenkov radiation emitted in these interactions is captured by an array of optical sensors embedded in the ice, which are known as Digital Optical Modules (DOMs). The neutrino events are reconstructed using the pattern of Cherenkov photons deposited inside the detector, using the number of photons as a proxy for the deposited energy and the timing information to reconstruct the neutrino trajectory direction.
 The events detected can be classified into different morphologies depending on the pattern of light deposition.
 \begin{itemize}
     \item \textbf{Cascades} are shower-like events produced via charged-current (CC) interaction of electron neutrinos ($\nu_e$) and neutral-current (NC) interactions of all three neutrino flavours. Tau neutrinos with a short decay length also contribute to this channel.
     \item \textbf{Tracks} are generated when muon neutrinos ($\nu_\mu$) undergo CC interactions, with Cherenkov photons emitted along a straight line by the outgoing muon.
     \item \textbf{Double cascades} are a  distinct feature of tau neutrino ($\nu_\tau$) CC interactions, where the tau lepton created also decays inside the detector volume. Both the tau creation and decay vertices observed as cascades.
 \end{itemize}
 A special class of IceCube events are named starting events. These are events where the interaction vertex of the neutrino is contained within the detector volume and a majority of the initial hadronic interaction is visible to the DOMs. The advantages of such 'starting event' samples are that they are sensitive to all neutrino flavours, and to events from all directions.\\
 IceCube is uniquely suited to the detection and measurement of the spectrum of high energy astrophysical neutrinos. These measurements were first reported in 2013~\cite{IceCube:2013_HE,PRL_PeV,IceCube:2014_3HE}.
 These neutrinos are theorized to be produced at sites of cosmic ray acceleration in the Universe. When high energy cosmic rays interact with matter or radiation in the ambient environment around these accelerators, or during their propagation through space, they create charged mesons which decay to yield neutrinos. Measurements of features in the all-sky neutrino energy spectrum are particularly interesting as they can reveal the underlying dynamics of a source population~\cite{murase_hidden_2016}, as well as shed light on the physics of cosmic ray acceleration.
 \section{Diffuse Astrophysical Neutrino Spectral Measurements with MESE}
 IceCube has performed many measurements of the diffuse neutrino spectrum, most recently a measurement using muon neutrino starting track events~\cite{ESTES}. These measurements all favored a single power law (SPL) astrophysical flux model as the best fit to data, although some deviations from the SPL were noted in a measurement using cascade events, electrons and tau neutrinos ~\cite{SBUCascades}.  \\
 We focus here on the measurement of the astrophysical neutrino flux with the Medium Energy Starting Events (MESE) sample~\cite{MESE_ICRC}. This is a sample sensitive to all flavors of neutrinos from the entire sky. This makes it ideal to resolve the diffuse neutrino spectrum, and test various flux hypotheses down to energies around 1~TeV. The primary background in these measurements is the large flux of atmospheric muons created in cosmic ray air showers. To gain sensitivity to the subdominant astrophysical neutrino flux, we must devise techniques to reject the atmospheric muon flux while retaining our neutrino signal.
Created to lower the sensitivity threshold of the flagship High Energy Starting Event (HESE) sample~\cite{IceCube:2013_HE,PRL_PeV,IceCube:2014_3HE,HESE7.5}  from 60~TeV to 1~TeV, MESE relies on a series of veto cuts to eliminate muon events, which are overwhelmingly likely to appear as track events originating outside the detector volume. The current iteration of the analysis is developed from a prior version~\cite{MESE_2yr}, with updated selection cuts and a new treatment of systematic uncertainties, in addition to more data. The stages of the MESE event selection are as follows.
 \begin{itemize}
     \item The \textbf{Outer Layer Veto}: Using the outermost layer of strings and DOM modules as a peripheral veto layer, we reject any event which deposits Cherenkov light in the veto layer before the interior of the detector. As muons tend to deposit light throughout their trajectory, this is an efficient way to reject atmospheric background
     \item The \textbf{Downgoing Track Veto}: The Outer Layer Veto is less efficient for low energy muons (around 1-50 TeV) , which deposit less light and often sneak through the veto layer before stochastically depositing light in the interior of the detector. To reject these muons, we search for isolated DOM photon hits associated with multiple downgoing track hypotheses passing through the vertex. If the number of hits is greater than a certain threshold, the event is likely a muon and is vetoed.
     \item The \textbf{Fiducial Volume Cut}: To reject the most pernicious background of low energy muons, we apply a charge- and zenith-dependent fiducial volume scaling. This essentially results in a larger veto region for the events which deposit the least light, efficiently reducing our muon background. 
 \end{itemize}
As MESE contains both cascade and track events, we apply a deep neural net (DNN) classifier~\cite{DNN_TheoGlauch} to separate events into the two morphologies, which are then reconstructed with different algorithms. Further details regarding the MESE event selection are available in~\cite{MESE_ICRC}.\\
\section{Analysis Results}\label{Analysis}
We apply a forward-folding binned likelihood analysis to measure the astrophysical neutrino spectrum, using various flux models. The likelihood of a given model is maximized by adjusting parameters that reweight the simulation to better describe the data. This is a computationally intensive multi-dimensional  operation, due to the large statistics simulation dataset used, and the multiple parameters that affect the predicted flux. The computation was performed using the \textsc{NNMFit}~\cite{NNMFit} software package, which uses the aesara~\cite{aesara} library to efficiently handle large tensor operations. \\
Various models were tested for the astrophysical flux. For each model, in addition to the physics parameters, we fit for various nuisance parameters that account for the different systematics uncertainties that contribute to the total measured event rate. In addition to the SPL, we test other spectral shapes, such as an exponential cutoff spectrum (SPE), a contribution from PeV-scale neutrinos from AGN cores \cite{SteckerAGN}, and a contribution from neutrinos from BL Lacertae objects \cite{BLLac}. The broken power law (BPL) and log-parabolic (LP) flux models add curvature to the spectrum across different energy ranges. Another model we tested includes  a Gaussian bump in the SPL in order to account for a potential excess at 30 TeV observed by the 2 year MESE result~\cite{MESE_2yr}.  In addition to these fits, we performed a model-independent fit of the astrophysical normalization in 13 energy segments assuming a power law spectrum in each segment.\\
The full list of tested models are shown in Tab. \ref{tab:results}, in order of decreasing log likelihood.

\begin{figure*}[tbh!]
    \begin{minipage}[b]{0.49\linewidth}
     \centering
        \includegraphics[width=\linewidth]{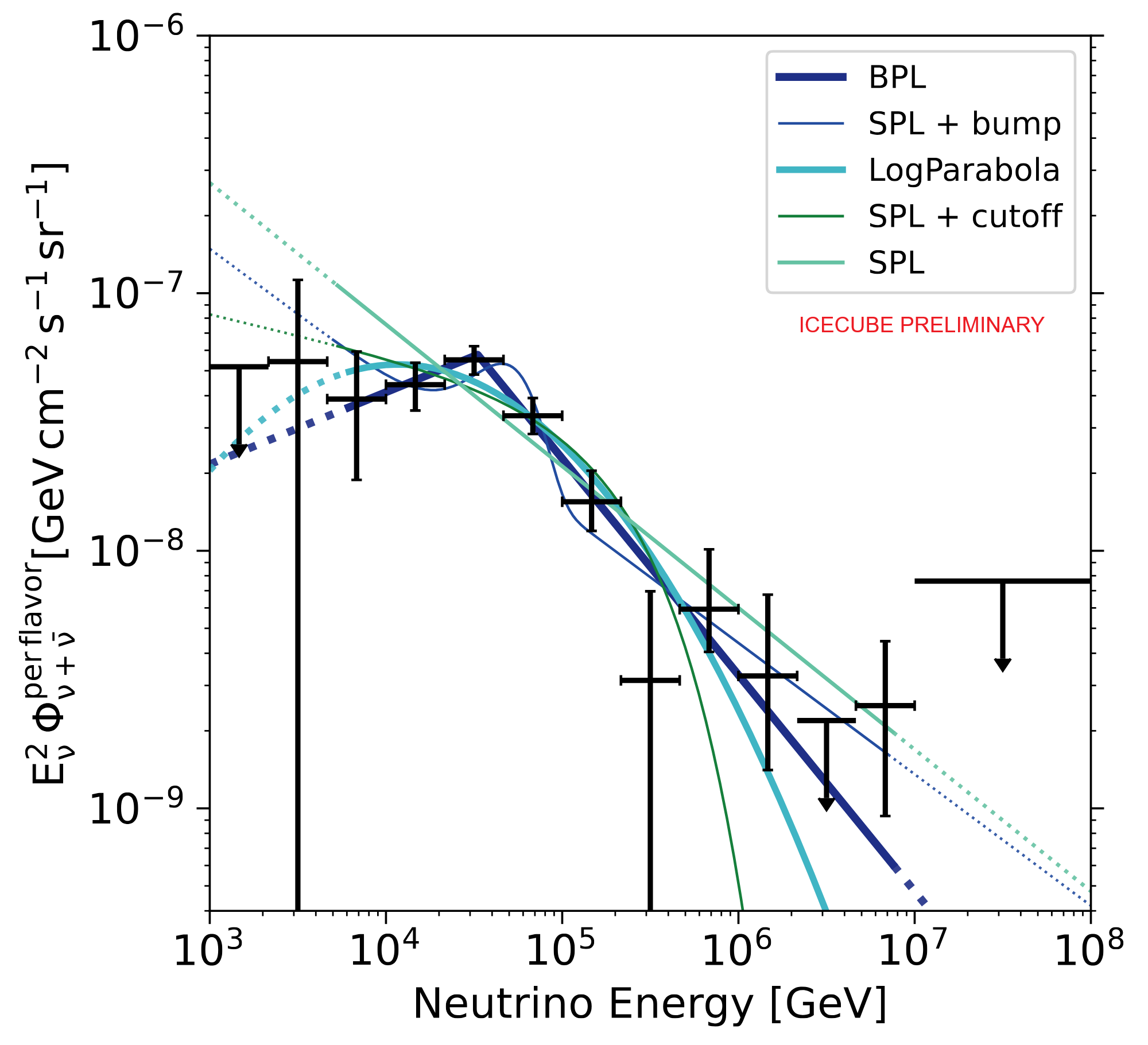}  
    \end{minipage}
    \begin{minipage}[b]{0.49\linewidth}
     \centering
        \includegraphics[width=\linewidth]{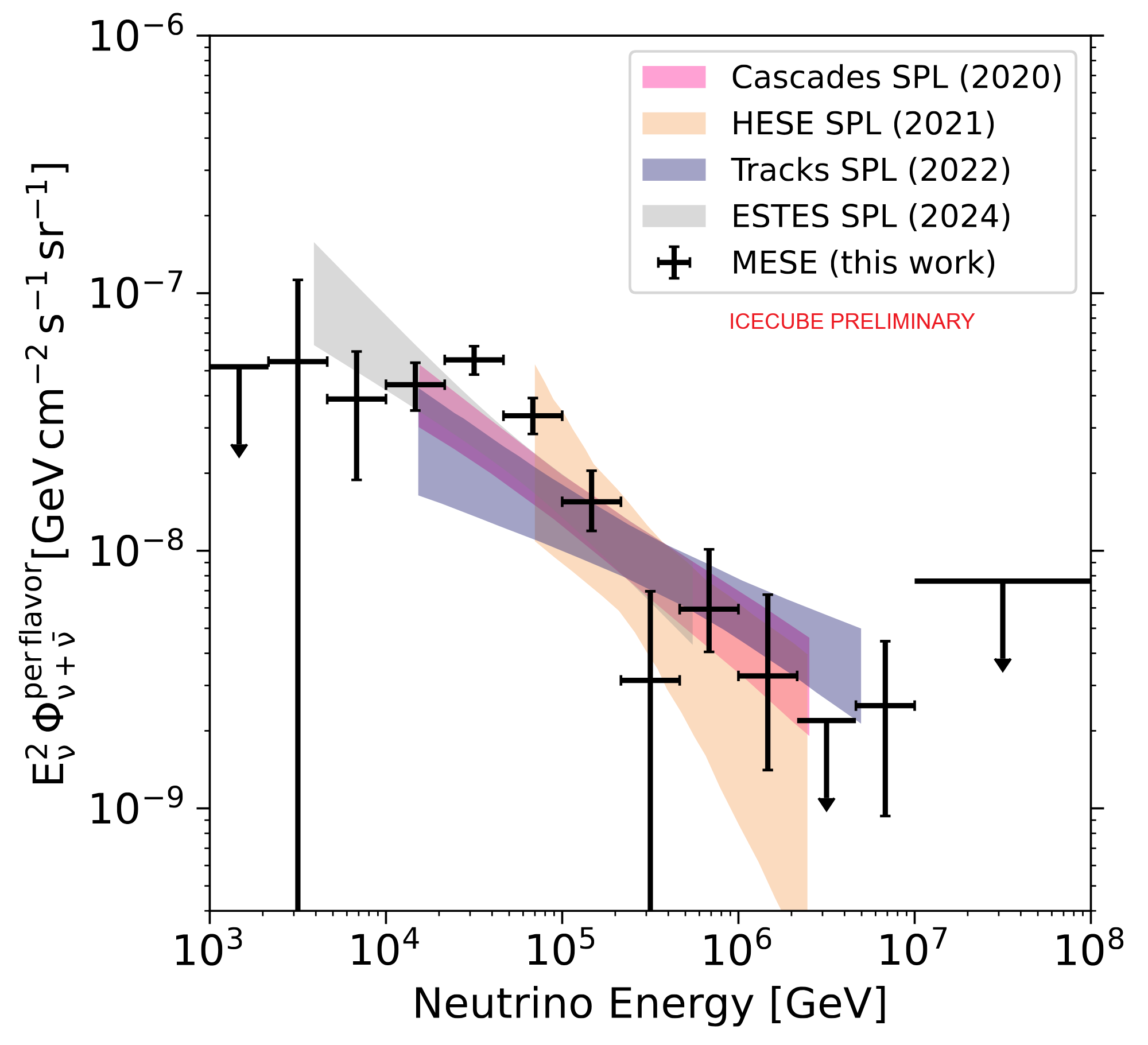}  
    \end{minipage}
\caption{\textbf{Model Fits to Data:} Results of a fit of the astrophysical neutrino flux in independent energy bands. \textit{Left:} The results are compared to the models fitted in the analysis. The solid lines show the energy range where the dataset is sensitive to the respective model, and the dotted lines show the energy range over which the fit is performed. \textit{Right:} The segmented fit is compared to previous measurements from IceCube, all favoring the SPL model. The shaded region corresponds to the 68\% confidence region for these measurements, as do the error bars on the MESE points
}
\label{fig:MESE_seg}
\end{figure*}
\begin{figure*}[t!]
    \begin{minipage}[b]{0.49\linewidth}
     \centering
        \includegraphics[width=\linewidth]{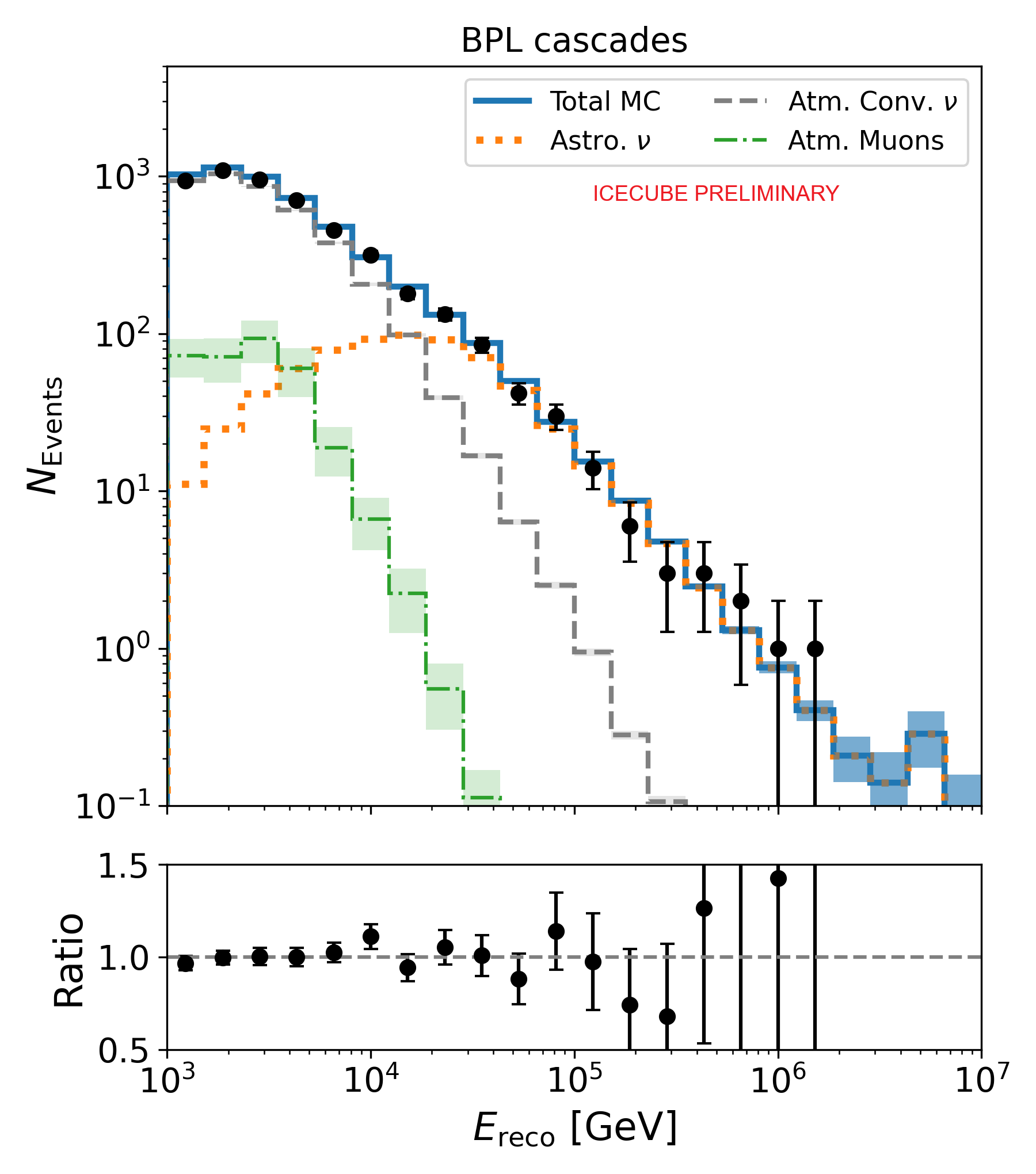}  
    \end{minipage}
    \begin{minipage}[b]{0.49\linewidth}
     \centering
        \includegraphics[width=\linewidth]{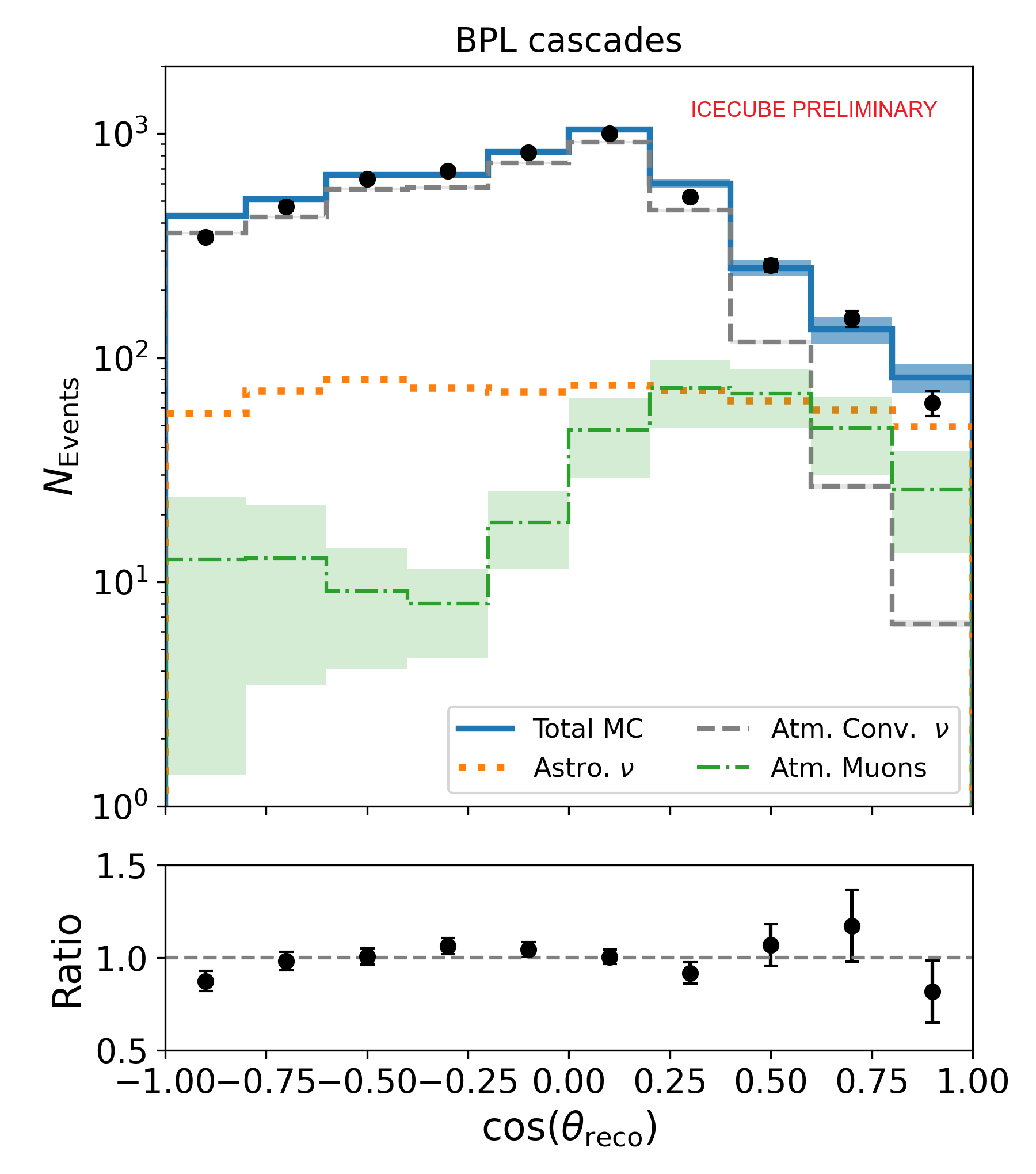}  
    \end{minipage}
     \begin{minipage}[b]{0.49\linewidth}
     \centering
        \includegraphics[width=\linewidth]{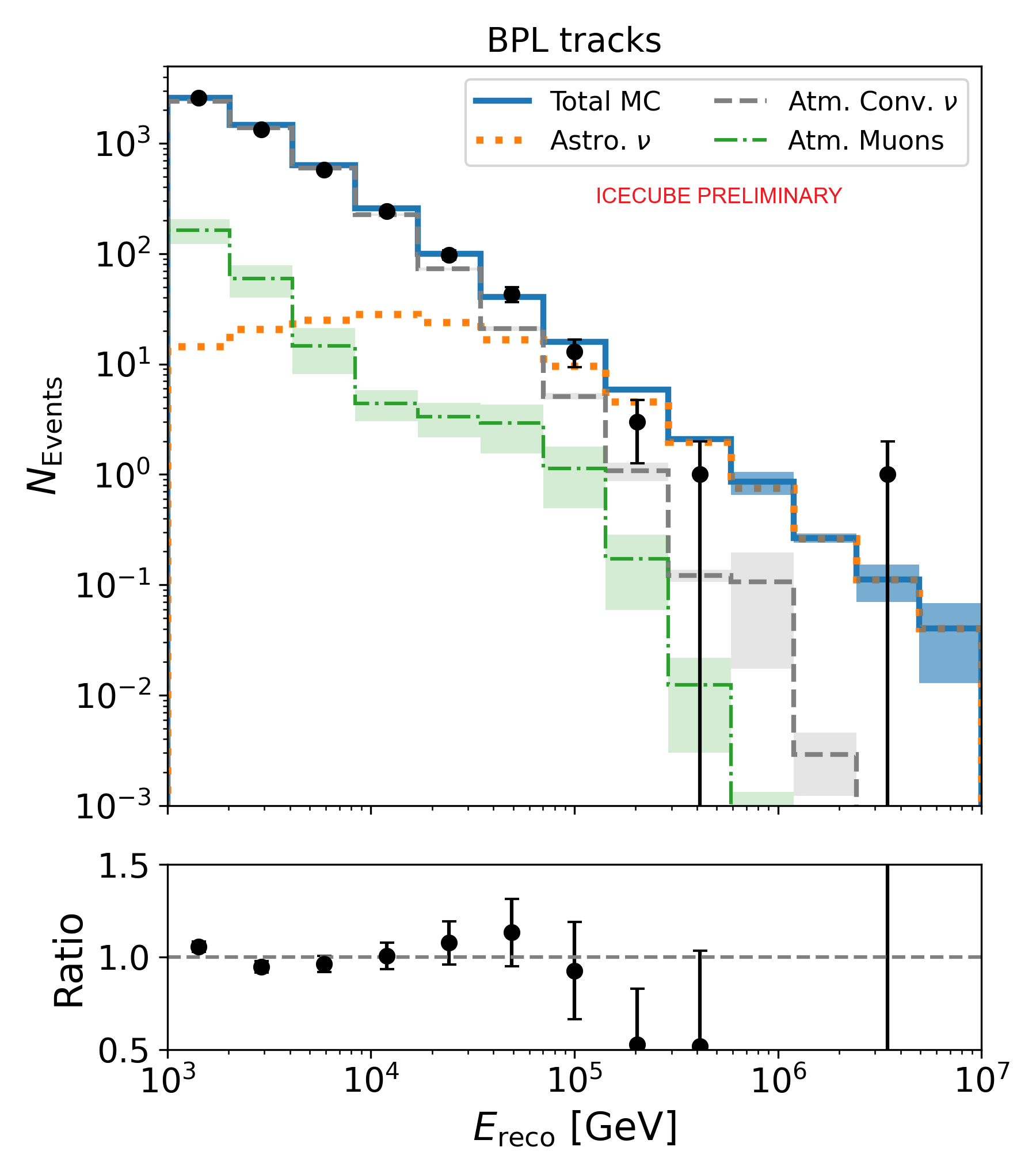}  
    \end{minipage}
    \begin{minipage}[b]{0.49\linewidth}
     \centering
        \includegraphics[width=\linewidth]{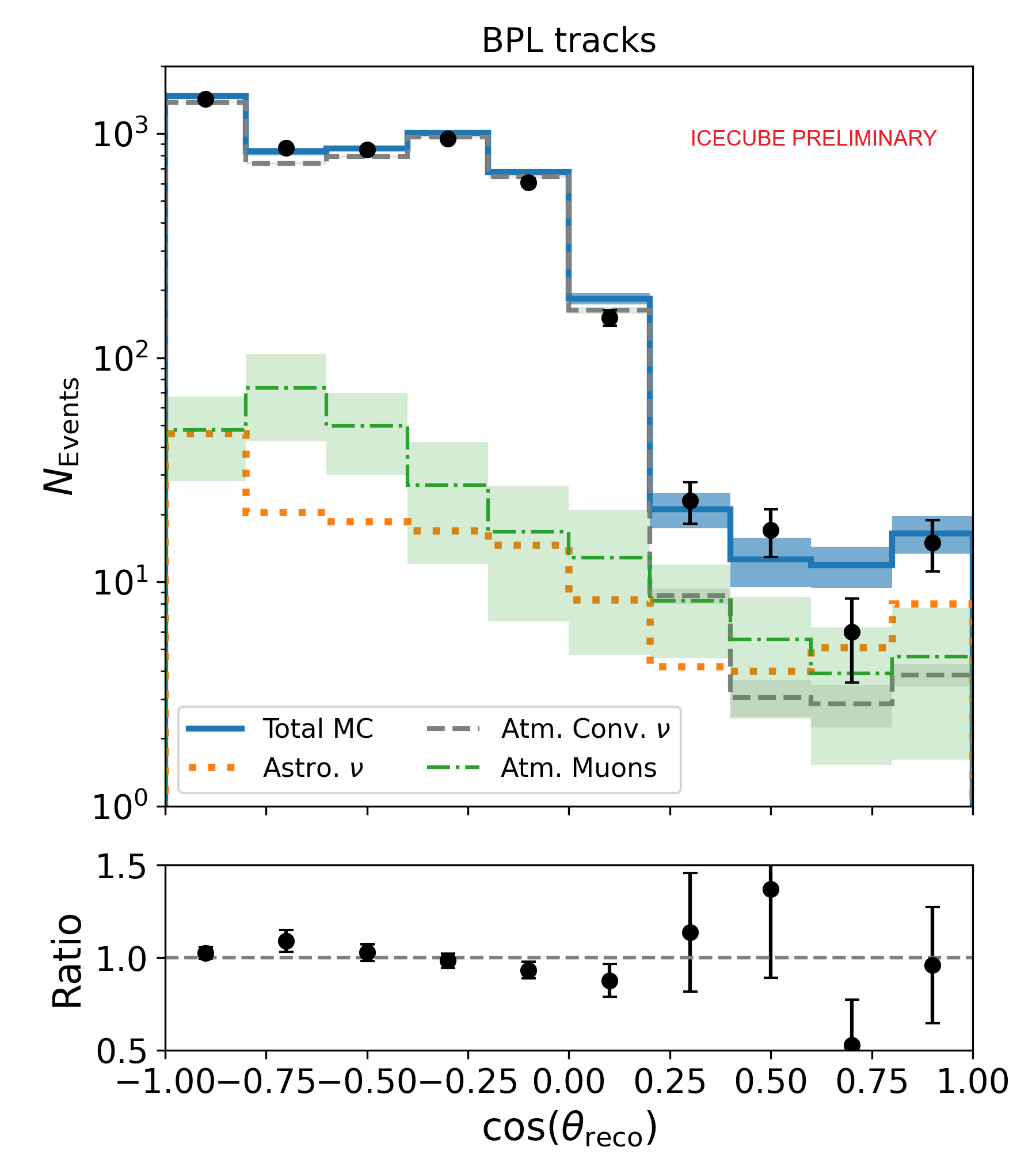}  
    \end{minipage}
\caption{\textbf{Comparison of data and simulation for the best fit BPL spectral model}: Displayed are reconstructed cascade energy (upper left), cos($\theta_{\rm{reco}}$)  (lower left), reconstructed track energy (upper right), and cos($\theta_{\rm{reco}}$) (lower right).  The atmospheric prompt neutrino flux normalization is a free parameter which fits to zero and is therefore omitted. The bottom panel of each plot shows the ratio of data and simulation in each bin.
}
\label{fig:MESEBPLDataMC}
\end{figure*}
\begin{table*}[htbp]
\caption[MESE Spectral Model Fits]{Results for the spectral models tested in the diffuse analysis. The uncertainties are derived from 1D profile likelihood scans, assuming Wilks's theorem applies. We show the preference over the single power-law hypothesis in terms of $-2 \Delta \mathrm{ln} \mathcal{L}$. The flux normalization is quoted per-flavor in units of $C=\rm{10^{-18}/GeV/cm^2/s/sr}$. All flux normalizations are at 100~TeV.}
\label{tab:results}
\renewcommand{\arraystretch}{1.5}
\centering
\small
\begin{tabularx}{\textwidth}{|c|c|c|}
\cline{1-3}
 Flux Model & Fit Parameters & \makecell[tc]{$-2 \Delta \mathrm{ln} \mathcal{L}$\\ (compared to SPL)} \\ \cline{1-3} 
\makecell[tc]{Single Power Law (SPL)\\$\left[
\Phi^{\nu+\bar{\nu}} (\frac{\mathrm{E}_{\nu}}{100 \mathrm{TeV}})^{-\gamma}\right]$ } & 
\begin{tabular}[t]{@{} l l @{}}
    $\Phi^{\nu+\bar{\nu}} / C$ & $= 2.13^{+0.18}_{-0.17}$ \\
    $\gamma$ & $= 2.55^{+0.04}_{-0.04}$
\end{tabular} & 
\begin{tabular}[t]{@{} l l @{}}
 & $0$
\end{tabular} \\ \cline{1-3} 
\makecell[tc]{SPL + AGN\\$\left[
\Phi^{\nu+\bar{\nu}} (\frac{\mathrm{E}_{\nu}}{100 \mathrm{TeV}})^{-\gamma} +\Phi_{\mathrm{model}} \right]$ } & 
\begin{tabular}[t]{@{} l l @{}}
    $\Phi^{\nu+\bar{\nu}} / C$ & $= 2.13^{+0.18}_{-0.17}$ \\
    $\gamma$ & $= 2.55^{+0.04}_{-0.04}$\\
   $\Phi_{\mathrm{model}}$& $=0^{+0.002}$
\end{tabular} & 
\begin{tabular}[t]{@{} l l @{}}
 & $0$
\end{tabular} \\ \cline{1-3} 
\makecell[tc]{SPL + BLLac\\$\left[
\Phi^{\nu+\bar{\nu}} (\frac{\mathrm{E}_{\nu}}{100 \mathrm{TeV}})^{-\gamma}\right]$ } & 
\begin{tabular}[t]{@{} l l @{}}
    $\Phi^{\nu+\bar{\nu}} / C$ & $= 2.13^{+0.18}_{-0.17}$ \\
    $\gamma$ & $= 2.55^{+0.04}_{-0.04}$\\
    $\Phi_{\mathrm{model}}$& $=0^{+0.002}$
\end{tabular} & 
\begin{tabular}[t]{@{} l l @{}}
 & $0$
\end{tabular} \\ \cline{1-3} 
\makecell[tc]{SPL + Cutoff\\$\left[
\Phi^{\nu+\bar{\nu}} (\Lambda)^{-\gamma} e^{\frac{-\mathrm{E}_{\nu}}{\mathrm{E}_\mathrm{cutoff}}}\right]$\\
$\Lambda=\frac{\mathrm{E}_{\nu}}{100 \mathrm{TeV}}$} & 
\begin{tabular}[t]{@{} l l @{}}
    $\Phi^{\nu+\bar{\nu}} / C$ & $= 3.975^{+1.14}_{-1.32}$ \\
    $\gamma$ & $= 2.16^{+0.23}_{-0.16}$ \\
    $\log_{10}(\frac{\mathrm{E}_\mathrm{cutoff}}{\mathrm{GeV}})$ & $= 5.40^{+0.51}_{-0.23}$ \\
\end{tabular} & 
\begin{tabular}[t]{@{} l l @{}}
 & \makecell[tc]{$1.8$\\$p= 0.18$\\ ($0.9\,\sigma$)}
\end{tabular} \\ \cline{1-3} 
\makecell[tc]{Log Parabola\\$\left[
\Phi^{\nu+\bar{\nu}} (\frac{\mathrm{E}_{\nu}}{100 \mathrm{TeV}})^{-\alpha_\mathrm{LP}-\beta_\mathrm{LP}\log_{10}(\frac{\mathrm{E}_{\nu}}{100 \mathrm{TeV}})
}\right]$} & 
\begin{tabular}[t]{@{} l l @{}}
    $\Phi^{\nu+\bar{\nu}} / C$ & $= 2.58^{+0.26}_{-0.26}$ \\
    $\alpha_\mathrm{LP}$ & $= 2.67^{+0.13}_{-0.06}$ \\
    $\beta_\mathrm{LP}$ & $= 0.36^{+0.10}_{-0.08}$ \\
\end{tabular} &
\begin{tabular}[t]{@{} l l @{}}
 & \makecell[tc]{$18.84$\\$p= 1.42\cdot 10^{-5}  $\\  ($4.2\,\sigma$)}
\end{tabular}\\ \cline{1-3} 
 \makecell[tc]{SPL + Bump\\$\left[
 \Phi^{\nu+\bar{\nu}} (\frac{\mathrm{E}_{\nu}}{100 \mathrm{TeV}})^{-\gamma} + \Phi_{\mathrm{bump}} e^{\frac{-(\mathrm{E}_{\nu}-\mathrm{E}_\mathrm{bump})^2}{2\sigma^2_{\mathrm{bump}}}}\right]$
} & 
\begin{tabular}[t]{@{} l l @{}}
    $\Phi^{\nu+\bar{\nu}}$ \, / \,C \, &$=1.42^{+0.21}_{-0.20}$ \\
    $\gamma$ & $=2.51^{+0.05}_{-0.07}$ \\
    $\log_{10}(\frac{\mathrm{E}_\mathrm{bump}}{\mathrm{GeV}})$ & $= 4.30^{+0.13}_{}$ \\
    $\log_{10}(\frac{\sigma_{\mathrm{bump}}}{\mathrm{GeV}})$ & $= 4.42^{+0.12}_{-0.13}$ \\
     $\Phi_{\mathrm{bump}}$ \, / \,C \, &$=24.79^{+13.55}_{-7.95}$ \\
\end{tabular} & 
\begin{tabular}[t]{@{} l l @{}}
 & \makecell[tc]{$22.3$\\$p= 5.65\cdot 10^{-5}$ \\$(3.9\, \sigma)$}
\end{tabular} \\ \cline{1-3} 
\makecell[tc]{Broken Power Law\\$\left[
\Phi^{\nu+\bar{\nu}}(\frac{\mathrm{E}_{\nu}}{\mathrm{E}_{\mathrm{break}}})^{-\gamma_{\mathrm{BPL}}}(\frac{\mathrm{E}_{\mathrm{break}}}{100 \mathrm{TeV}})^{-\gamma_1}\right]$\\
$\gamma_{\mathrm{BPL}}=\left\{
\begin{array}{c}
       \gamma_1\, (\mathrm{E}_{\nu} < E_{\mathrm{break}})\\
       \gamma_2\, (\mathrm{E}_{\nu} > E_{\mathrm{break}})
    \end{array}\right.$} & 
\begin{tabular}[t]{@{} l l @{}}
    $\Phi^{\nu+\bar{\nu}} / C$ & $= 2.28^{+0.22}_{-0.20}$ \\
    $\gamma_1$ & $= 1.72^{+0.26}_{-0.35}$ \\
    $\gamma_2$ & $= 2.84^{+0.11}_{-0.09}$ \\
    $\log_{10}(\frac{\mathrm{E}_\mathrm{break}}{\mathrm{GeV}})$ & $= 4.52^{+0.11}_{-0.09}$ \\
\end{tabular} & 
\begin{tabular}[t]{@{} l l @{}}
 & \makecell[tc]{$27.3$\\
 $p= 1.2\cdot 10^{-6}  $ \\ ($4.7\,\sigma$)}
\end{tabular} \\ \cline{1-3} 
\end{tabularx}
\end{table*}
\section{Discussion}\label{Discussion}
We see that the best fit to the data is provided by the broken power law spectral model, with the test statistic $\mathrm{TS}\,=\,-2 \Delta \mathrm{ln} \mathcal{L}\,=\,27.3$ when compared to the likelihood obtained for the SPL model. This corresponds to a p-value of $1.2\times10^{-6}$, or $4.7\,\sigma$, assuming Wilks's Theorem~\cite{Wilks}. Figure~\ref{fig:MESEBPLDataMC} shows the agreement between the data and the modeling of the various contributions to the neutrino flux. This result, in conjunction with a contemporaneous measurement in \cite{naab2023measurementastrophysicaldiffuseneutrino}, marks the first time IceCube has been able to resolve structure in the diffuse neutrino spectrum beyond a single power law to a degree above $4\,\sigma$. As evident in Fig.~\ref{fig:MESE_seg}, two features appear to drive this preference for curvature in the spectrum. These are the excess at $\sim\,30$ TeV and a deficit at a few hundred TeV, when compared to the baseline SPL model.\\
Various tests were performed to validate the results obtained by this analysis, both before and after unblinding the data. In particular, it was shown that any non-isotropic neutrino flux from the galactic plane~\cite{GalacticPlaneScience} failed to bias the best fit physics parameters to a significant degree. Similarly, possible variations in the modeling of the atmospheric neutrino flux or a contribution from `prompt' neutrinos from charmed meson decay in the atmosphere did not change the physics parameters significantly. Full details of the robustness checks performed are provided in \cite{basu2025characterization}.\\
The observation of curvature in the neutrino spectrum is a departure from the results of IceCube's previous analyses, as illustrated in Fig.~\ref{fig:MESE_seg}. The most recent ESTES measurement~\cite{ESTES} is similarly sensitive down to $\mathcal{O}$(1 TeV), and shows no preference for a departure from the single power law. It is difficult to directly compare the ESTES and MESE analyses as both use different simulation datasets, and a different treatment of systematic uncertainties. A particular point of difference is the treatment of the atmospheric neutrino self-veto~\cite{SelfVeto_AK_JVS,nuVeto}. It is worth noting that a segmented fit performed using MESE's tracks subset, which also contains primarily starting tracks, is compatible with the ESTES measurement, and it is clear that further investigation is necessary to resolve the differences in the various samples. 
\section{Conclusion}\label{Conclusion}
\begin{justify}
We present the results of the measurement of the diffuse astrophysical neutrino spectrum using the MESE event selection. Using updated veto techniques to reject the atmospheric muon background, MESE is sensitive to the astrophysical neutrino flux down to $\mathcal{O}$(1 TeV). With this sensitivity, we find a departure from the single power law, with a harder flux below $\sim30$ TeV, and a steeper flux at higher energies.\\
This result agrees with a contemporary combined measurement from IceCube's cascades and upgoing tracks samples~\cite{RNaabThesis,EGansterThesis,naab2023measurementastrophysicaldiffuseneutrino}, and is the first time IceCube has been able to resolve structure in the cosmic neutrino spectrum beyond $4\,\sigma$ significance. A joint publication describing these analyses is in preparation, and will describe the results more fully, along with the verification checks performed.\\
The departure from IceCube's prior single power law measurements, most recently the ESTES analysis, requires further study. Efforts are currently underway to unify these channels into a joint measurement, using the same treatment of systematic uncertainties and a common analysis framework. The next generation of diffuse measurements will shed light on the features of the astrophysical neutrino spectrum, and help uncover the dynamics of cosmic ray accelerators in the Universe.
\end{justify}
\bibliographystyle{ICRC}
\bibliography{references}

%

\newpage
\section*{Full Author List: IceCube Collaboration}

\scriptsize
\noindent
R. Abbasi$^{16}$,
M. Ackermann$^{63}$,
J. Adams$^{17}$,
S. K. Agarwalla$^{39,\: {\rm a}}$,
J. A. Aguilar$^{10}$,
M. Ahlers$^{21}$,
J.M. Alameddine$^{22}$,
S. Ali$^{35}$,
N. M. Amin$^{43}$,
K. Andeen$^{41}$
% ...
\\
\\
% ...
$^{10}$ Universit{\'e} Libre de Bruxelles, Science Faculty CP230, B-1050 Brussels, Belgium \\
% ...
$^{16}$ Department of Physics, Loyola University Chicago, Chicago, IL 60660, USA \\
$^{17}$ Dept. of Physics and Astronomy, University of Canterbury, Private Bag 4800, Christchurch, New Zealand \\
% ...
$^{21}$ Niels Bohr Institute, University of Copenhagen, DK-2100 Copenhagen, Denmark \\
$^{22}$ Dept. of Physics, TU Dortmund University, D-44221 Dortmund, Germany \\
% ...
$^{35}$ Dept. of Physics and Astronomy, University of Kansas, Lawrence, KS 66045, USA \\
% ...
$^{39}$ Dept. of Physics and Wisconsin IceCube Particle Astrophysics Center, University of Wisconsin{\textemdash}Madison, Madison, WI 53706, USA \\
% ...
$^{41}$ Department of Physics, Marquette University, Milwaukee, WI 53201, USA \\
% ...
$^{43}$ Bartol Research Institute and Dept. of Physics and Astronomy, University of Delaware, Newark, DE 19716, USA \\
% ...
$^{63}$ Deutsches Elektronen-Synchrotron DESY, Platanenallee 6, D-15738 Zeuthen, Germany \\
$^{\rm a}$ also at Institute of Physics, Sachivalaya Marg, Sainik School Post, Bhubaneswar 751005, India \\
% ...

\subsection*{Acknowledgments}

\noindent
The authors gratefully acknowledge the support from the following agencies and institutions:
USA {\textendash} U.S. National Science Foundation-Office of Polar Programs,
U.S. National Science Foundation-Physics Division,
U.S. National Science Foundation-EPSCoR,
U.S. National Science Foundation-Office of Advanced Cyberinfrastructure,
Wisconsin Alumni Research Foundation,
Center for High Throughput Computing (CHTC) at the University of Wisconsin{\textendash}Madison,
Open Science Grid (OSG),
Partnership to Advance Throughput Computing (PATh),
Advanced Cyberinfrastructure Coordination Ecosystem: Services {\&} Support (ACCESS),
Frontera and Ranch computing project at the Texas Advanced Computing Center,
U.S. Department of Energy-National Energy Research Scientific Computing Center,
Particle astrophysics research computing center at the University of Maryland,
Institute for Cyber-Enabled Research at Michigan State University,
Astroparticle physics computational facility at Marquette University,
NVIDIA Corporation,
and Google Cloud Platform;
Belgium {\textendash} Funds for Scientific Research (FRS-FNRS and FWO),
FWO Odysseus and Big Science programmes,
and Belgian Federal Science Policy Office (Belspo);
Germany {\textendash} Bundesministerium f{\"u}r Forschung, Technologie und Raumfahrt (BMFTR),
Deutsche Forschungsgemeinschaft (DFG),
Helmholtz Alliance for Astroparticle Physics (HAP),
Initiative and Networking Fund of the Helmholtz Association,
Deutsches Elektronen Synchrotron (DESY),
and High Performance Computing cluster of the RWTH Aachen;
Sweden {\textendash} Swedish Research Council,
Swedish Polar Research Secretariat,
Swedish National Infrastructure for Computing (SNIC),
and Knut and Alice Wallenberg Foundation;
European Union {\textendash} EGI Advanced Computing for research;
Australia {\textendash} Australian Research Council;
Canada {\textendash} Natural Sciences and Engineering Research Council of Canada,
Calcul Qu{\'e}bec, Compute Ontario, Canada Foundation for Innovation, WestGrid, and Digital Research Alliance of Canada;
Denmark {\textendash} Villum Fonden, Carlsberg Foundation, and European Commission;
New Zealand {\textendash} Marsden Fund;
Japan {\textendash} Japan Society for Promotion of Science (JSPS)
and Institute for Global Prominent Research (IGPR) of Chiba University;
Korea {\textendash} National Research Foundation of Korea (NRF);
Switzerland {\textendash} Swiss National Science Foundation (SNSF).

\end{document}